\documentclass{article}

\usepackage[preprint]{neurips_2025}
    
\usepackage[utf8]{inputenc} 
\usepackage[T1]{fontenc}    
\usepackage{hyperref}       
\usepackage{url}            
\usepackage{booktabs}       
\usepackage{amsfonts}       
\usepackage{nicefrac}       
\usepackage{microtype}      
\usepackage{xcolor}         
\usepackage{amsmath}
\usepackage{amsthm}
\newtheorem{prop}{Proposition}
 \usepackage{graphicx}

\newcommand{\beginappendix}{
            \newpage
            \renewcommand{\thesection}{A}
            \setcounter{table}{0}
            \renewcommand{\thetable}{A\arabic{table}}
            \setcounter{figure}{0}
            \renewcommand{\thefigure}{A\arabic{figure}}
            \setcounter{equation}{0}
            \renewcommand{\theequation}{A\arabic{equation}}
            }

\title
{SPINT: Spatial Permutation-Invariant Neural Transformer for Consistent Intracortical Motor Decoding}

\author{
  Trung Le\\
  \small{University of Washington}\\
  \small\texttt{tle45@uw.edu} \\
  \And
  Hao Fang\\
  \small{University of Washington}\\
  \small\texttt{haof459@uw.edu} \\
  \AND
  Jingyuan Li\\
  \small{University of Washington}\\
  \small\texttt{jingyli6@uw.edu} \\
  \And
  Tung Nguyen\\
  \small{University of California, Los Angeles}\\
  \small\texttt{tungnd@cs.ucla.edu} \\
  \And
  Lu Mi\\
  \small{Tsinghua University}\\
  \small\texttt{milu@mail.tsinghua.edu.cn} \\
  \And
  Amy Orsborn\\
  \small{University of Washington}\\
  \small\texttt{aorsborn@uw.edu} \\
  \And
  Uygar Sümbül\\
  \small{Allen Institute}\\
  \small\texttt{uygars@alleninstitute.org} \\
  \And
  Eli Shlizerman\\
  \small{University of Washington}\\
  \small\texttt{shlizee@uw.edu} \\
}

\begin{document}

\maketitle

\begin{abstract}
Intracortical Brain-Computer Interfaces (iBCI) aim to decode behavior from neural population activity, enabling individuals with motor impairments to regain motor functions and communication abilities. A key challenge in long-term iBCI is the nonstationarity of neural recordings, where the composition and tuning profiles of the recorded populations are unstable across recording sessions. Existing methods attempt to address this issue by explicit alignment techniques; however, they rely on fixed neural identities and require test-time labels or parameter updates, limiting their generalization across sessions and imposing additional computational burden during deployment. In this work, we introduce SPINT - a \textbf{S}patial \textbf{P}ermutation-\textbf{I}nvariant \textbf{N}eural \textbf{T}ransformer framework for behavioral decoding that operates directly on unordered sets of neural units. Central to our approach is a novel \emph{context-dependent positional embedding} scheme that dynamically infers unit-specific identities, enabling flexible generalization across recording sessions. SPINT supports inference on variable-size populations and allows few-shot, gradient-free adaptation using a small amount of unlabeled data from the test session. To further promote model robustness to population variability, we introduce \emph{dynamic channel dropout}, a regularization method for iBCI that simulates shifts in population composition during training. We evaluate SPINT on three multi-session datasets from the FALCON Benchmark, covering continuous motor decoding tasks in human and non-human primates. SPINT demonstrates robust cross-session generalization, outperforming existing zero-shot and few-shot unsupervised baselines while eliminating the need for test-time alignment and fine-tuning. Our work contributes an initial step toward a robust and scalable neural decoding framework for long-term iBCI applications.

\end{abstract}

\section{Introduction}
Motor behavior arises from the complex interplay between interconnected neurons, each possessing distinct functional properties \cite{georgopoulos1986neuronal}. Deciphering the highly nonlinear mapping from the activity of these neural populations to behavior has been a major focus of intracortical Brain-Computer Interfaces (iBCI), whose applications have enabled individuals with motor impairments to control external devices \cite{collinger2013high}, restore communication abilities through typing \cite{pandarinath2017high}, handwriting \cite{willett2021high}, and speech \cite{willett2023high}. 

Despite the remarkable capabilities, iBCI systems suffer from performance degradation over extended periods of time, largely attributed to the nonstationarities of the recorded populations \cite{chestek2011long}. Sources of nonstationarities include shifts in electrode position, tissue impedance changes, and neural plasticity \cite{gallego2020long, perge2013intra, downey2018intracortical}. These nonstationarities lead to changes in the number and identity of neural units picked up by recording electrodes over time. Such changes in population composition alter the learned neural activity to behavior mapping, preventing decoders trained on previous sessions to maintain robust performance on new sessions. 
To ensure robustness of behavior decoding over future recording sessions, one approach has focused on training deep networks using many sessions, attempting to achieve decoders that are robust to the cross-session variability \cite{sussillo2016making, hosman2023months}. This zero-shot approach requires months of labeled training data, necessitating extensive data collection from the user. While recent methods targeting cross-subject generalization may alleviate some of this burden \cite{rizzoglio2023monkeys, safaie2023preserved, ye2023neural, azabou2023unified}, degradation over long-term use still remains, advocating for the adoption of adaptive methods \cite{karpowicz2024few, dyer2017cryptography, degenhart2020stabilization, gallego2020long, karpowicz2022stabilizing}. These adaptive methods leverage the low-dimensional manifold underlying population activity that has been shown to preserve a consistent relationship with behavior over long periods of time \cite{pandarinath2018inferring, gallego2020long}. Depending on the use of labels at test time, they can be categorized into supervised \cite{ye2023neural, ye2025generalist, azabou2023unified}, semi-supervised \cite{fan2023plug}, or unsupervised \cite{karpowicz2022stabilizing, dyer2017cryptography, ma2023using}, with varying level of success and practical utility in real-world iBCI \cite{karpowicz2024few}.

Despite the variety of technical approaches, these works share a common design philosophy: they adopt a fixed view of the neural population, assigning fixed identities and order for neural units during training. While this treatment achieves high decoding performance on held-in sessions (within-session generalization), the decoders suffer from out-of-distribution performance degradation when evaluated on held-out sessions with different sizes and unit membership (cross-session generalization). To enable transfer of the pretrained model to novel sessions, explicit alignment procedures with gradient updates to adapt model parameters are necessary, imposing disruptive and costly computation overhead for iBCI users. With these limitations of the existing approaches, we advocate for the view that an ideal, universal iBCI decoder should be invariant to the permutation of the neural population by design, and should be able to seamlessly handle inference of a variable-sized, unordered set of neural units with minimal data collected from the new setting.

In this work, we introduce SPINT - a permutation-invariant framework that can decode motor behavior from the activity of unordered sets of neural units. We contribute toward an iBCI design that can predict behavior covariates from continuous streams of neural observations and adapt gradient-free to novel sessions with few-shot unlabeled calibration data. At the core of our methods is a permutation-invariant transformer with a novel context-dependent positional embedding that allows flexible identification of neural unit identities on-the-fly. We further introduce \emph{dynamic channel dropout}, a novel regularization method to encourage model robustness to neural population composition. We evaluate our approach on three movement decoding datasets from the FALCON Benchmark \cite{karpowicz2024few}, demonstrating robust cross-session generalization on motor tasks in human and non-human primates. Our model outperforms zero-shot and few-shot unsupervised baselines, while not requiring any retraining or fine-tuning overhead.   

In summary, the contributions of this work include:

\begin{itemize}
\item We present a \emph{transformer-based permutation-invariant framework} with a novel \emph{context-dependent positional embedding} for few-shot unsupervised behavioral decoding. Our flexible, lightweight model enables ingestion of unordered sets of neural units during training and facilitates out-of-the-box inference on unseen neural populations.
\item We introduce \emph{dynamic channel dropout}, a novel regularization technique for iBCI applications to promote decoder robustness to the composition of input neural population.
\item We evaluate our model on three motor behavioral decoding datasets in the FALCON Benchmark, showing robust gradient-free generalization to unseen sessions in the presence of cross-session nonstationarities.
\end{itemize}
\begin{figure}
    \centering
    \includegraphics[width=1.0\linewidth]{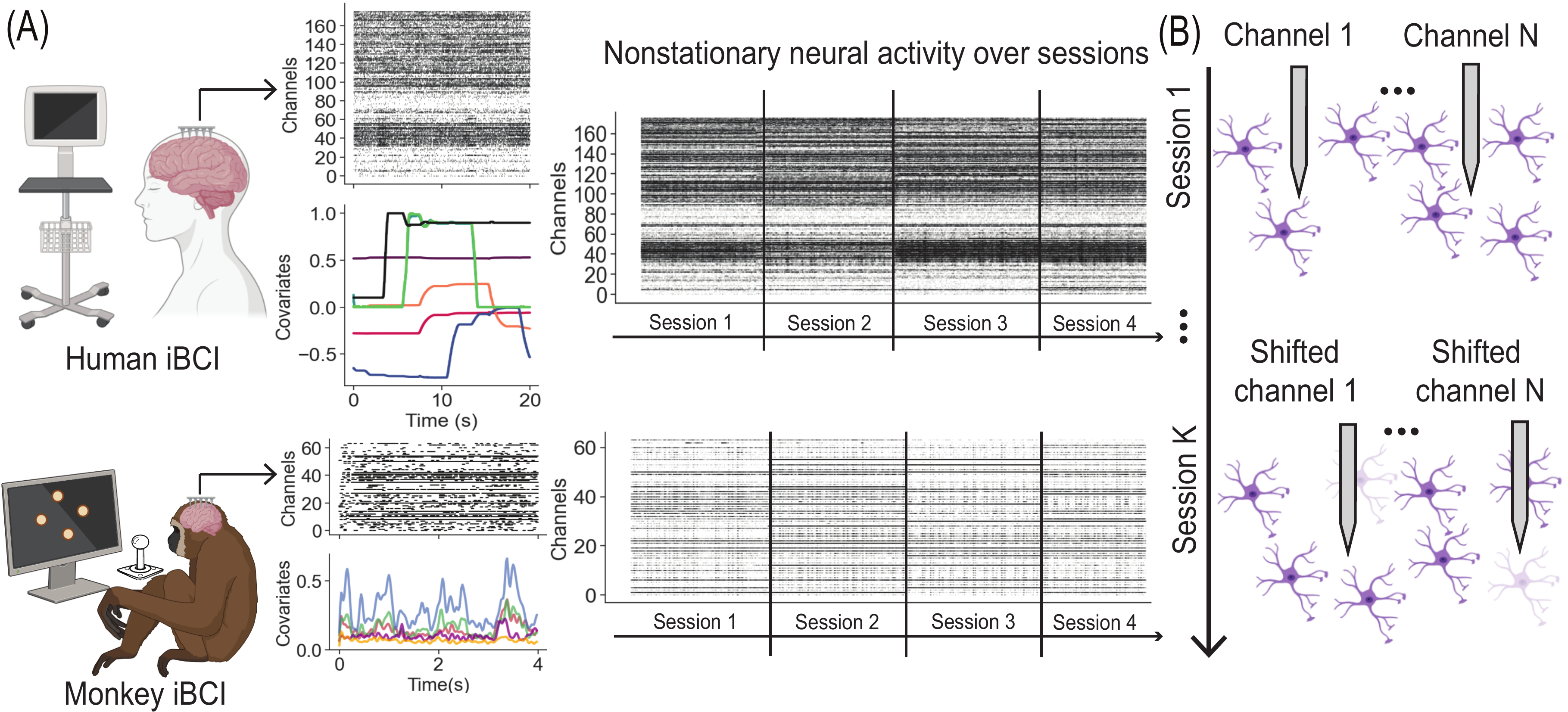}
    \caption{\textbf{Nonstationarities in long-term iBCI.} (A) Examples of iBCI systems in human and non-human primates. Spiking activity is recorded from multichannel electrode arrays together with behavior covariates, e.g., 7 degree-of-freedom robotic arm control or electromyography from the upper limb. Neural activity exhibits nonstationarities over recording sessions. (B) Systematic changes in neuron positions, including the introduction or loss of neurons in the vicinity of electrodes and the shifts of the entire electrode array can contribute to instability of neural recordings over time.}
    \label{fig 1: ibci background}
\end{figure}
\section{Related Work}
\textbf{Decoding motor behavior from intracortical population activity}: A variety of computational models have been proposed to model population spiking dynamics and decode motor behavior, starting with linear models such as population vectors \cite{georgopoulos1986neuronal}, linear regression \cite{paninski2004maximum}, Kalman filter \cite{wu2006bayesian, pistohl2008prediction, wu2009neural}. Nonlinear methods have also been developed; a non-exhaustive list includes generalized linear models \cite{truccolo2005point, yu2005extracting, pillow2008spatio, lawhern2010population}, latent variable models \cite{wu2017gaussian, yu2008gaussian, kim2021inferring, schneider2023learnable}, leveraging deep neural networks including recurrent neural networks (RNN) \cite{pandarinath2018inferring, lee2024identifying, saxena2022motor} and transformers \cite{ye2021representation, ye2023neural, ye2025generalist, liu2022seeing, le2022stndt, azabou2023unified, zhang2024towards, zhang2025neural}. Most of these works assume a fixed view of the population across time and have limited direct transferability to unseen populations. We adopt the view of neural population as an unordered set of neural units, offering more flexibility in transferring across population compositions.

\textbf{Cross-session neural alignment methods}: 
Neural nonstationarities pose challenges for maintaining decoder performance over long periods of time. To tackle this issue, alignment techniques were proposed to align the testing sessions to match the distribution of a training session, usually performed in the latent space. These techniques vary from linear methods using Canonical Correlation Analysis (CCA)  \cite{gallego2020long, safaie2023preserved}, linear stabilizer \cite{degenhart2020stabilization}, linear distribution alignment \cite{dyer2017cryptography}, to nonlinear using generative adversarial networks (GAN) \cite{farshchian2018adversarial, ma2023using}, RNN \cite{karpowicz2022stabilizing, fan2023plug, vermani2023leveraging}, and diffusion models \cite{wang2023extraction}. Recent advances in foundation models for intracortical iBCI promise transferability across multiple sessions \cite{ye2023neural, ye2025generalist, azabou2023unified, zhang2024towards, zhang2025neural}. All these approaches, however, still require explicit alignment procedures with model parameter updates and test-time labels in some cases to adapt the pretrained decoder to unseen populations. 
On the other hand, our proposed context-dependent positional embedding scheme allows flexible, gradient-free adaptation in new sessions without any labels.
 
\textbf{Permutation-invariant neural networks for set-structured inputs}: While conventional neural networks are designed for fixed dimensional data instances, in many set-structured applications such as point cloud object recognition or image tagging, the inputs have no intrinsic ordering, advocating for a class of models that are permutation-invariant by design \cite{zaheer2017deep, qi2017pointnet}. One such work, DeepSets, introduced a set average pooling approach serving as a universal approximator for any set function \cite{zaheer2017deep}. Follow-up works \cite{qi2017pointnet, lee2019set} extended this pooling method to include max-pooling and attention mechanisms \cite{vaswani2017attention}. We take inspiration from these works to design our universal neural unit identifier and permutation-invariant behavior decoding framework, leveraging the permutation-invariant property of the attention mechanism \cite{xu2024permutation}. 
\section{Approach}
\begin{figure}
    \centering
    \includegraphics[width=1.0\linewidth]{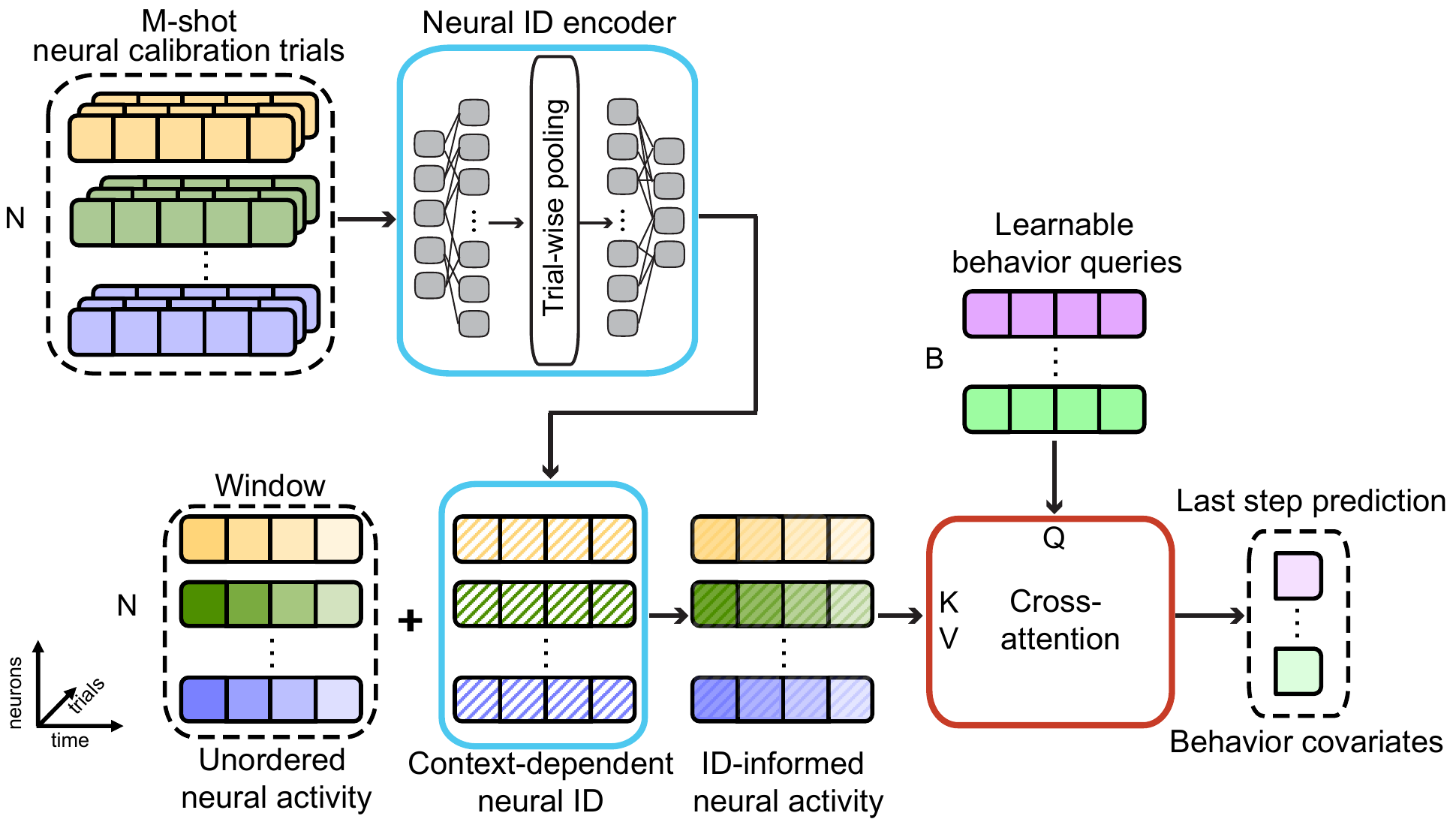}
    \caption{\textbf{SPINT architecture.} The model performs continuous behavioral decoding by predicting behavior covariates at the last timestep given a past window of activity from an unordered set of neural units. The universal Neural ID Encoder infers identities of the units using few-shot unlabeled calibration trials, while the cross-attention mechanism selectively aggregates information from the units to decode behavior.}
    \label{fig 2: algorithm}
\end{figure}
\subsection{A permutation-invariant framework for few-shot continuous behavioral decoding}
We study the problem of real-time, cross-session iBCI decoding, where behavior needs to be decoded in a causal manner from a continuous stream of neural observations. Concretely, within a single session $s$, let $X_{i,t}$ denote the binned spiking activity of neural unit $i$ at time $t$, $X_{i,:}$ denote all the activity of unit $i$, and $X_{:,t}$ denote the activities of all units at time $t$. Given a past observation window of population activity $X_{:, t-W+1:t} \in \mathbb{R}^{N_s \times W}$, where $N_s$ is the number of recorded neural units in session $s$ and $W$ is the length of the observation window, we aim to estimate the corresponding last time step of behavior output $Y_t \in \mathbb{R}^B$, where $B$ is the dimensionality of behavior covariates. Model parameters are fitted with gradient descent using labeled data from $k$ training (held-in) sessions and evaluated on $k'$ testing (held-out) sessions without gradient updates or labels. At our disposal on each held-out session is a short calibration period $X_{i,[C]} \in \mathbb{R}^{T'}$ consisting of $M$-shot variable-length trials lasting for $T'$ timesteps, to be used for cross-session adaptation.

Traditional approaches consider the population activity at each timestep as a "token" and decode behavior by modeling temporal dynamics of population activity \cite{yu2008gaussian, pandarinath2018inferring, ye2021representation}. By treating temporal snapshots of population activity $X_{:,t} \in \mathbb{R}^{N_s}$ as input vectors, these approaches assume a fixed number and order of neural units, requiring explicit spatial re-alignment when applied to another session with a different size and order \cite{pandarinath2018inferring, karpowicz2022stabilizing, ma2023using, safaie2023preserved, dyer2017cryptography}. Recent methods incorporating factorized spatial-temporal modeling \cite{le2022stndt, liu2022seeing} face similar challenges, while approaches with explicit spatiotemporal tokens \cite{ye2023neural, ye2025generalist, azabou2023unified} still require fine-tuning unit identity in novel sessions. These design choices hinder out-of-the-box generalizability of neural decoders across sessions, as a universal decoder should ideally be invariant to the permutation and size of the input population. 

To realize this goal, we treat windows of individual neural units $X_{i,t-W+1:t} \in \mathbb{R}^W$ as an \emph{unordered set} of tokens and aggregate information from these units to decode behavior using the cross-attention mechanism~\cite{vaswani2017attention}. To compensate for the loss of consistent order that the decoder can leverage for behavior decoding, we embed a notion of neural identity to each unit based on its spiking signature during a few-shot, \emph{unlabeled} calibration period $X_{:,[C]}$ in the same session. $X_{:,[C]}$ is either provided in limited amount in held-out sessions at test time, or is artificially sampled from held-in sessions during training. This context-dependent neural identity is inferred by a universal neural identity encoder that is shared across units and sessions, enabling \emph{gradient-free} adaptation to novel population compositions at test time. 
\subsection{Encoding identity of neural units}
Let $X_i^C \in \mathbb{R}^{M \times T}$ be the trialized version of $X_{i,[C]} \in \mathbb{R}^{T'}$. $X_i^C$ is the collection of $M$ calibration trials of neural unit $i$ interpolated to a fixed length $T$. We infer neural identity $E_i \in \mathbb{R}^{W}$ of unit $i$ by a neural network $\mathrm{IDEncoder}$:

\begin{equation}
    E_i = \mathrm{IDEncoder}(X_{i}^{C}) = \psi(pool(\phi(X_{i}^{C})))
\end{equation}

where $\psi$ and $\phi$ are multi-layer fully connected networks and $pool$ is the mean pooling operation across $M$ trials. Due to the permutation invariant nature of the mean operation and the fact that $\psi$ and $\phi$ are applied trial-wise, $\mathrm{IDEncoder}$ is invariant to the order of $M$ calibration trials by design \cite{zaheer2017deep}. 
\subsection{Decoding behavior via selective aggregation of information from neural population units}
After inferring the identity for each unit from its calibration period, we add $E_i$ to all $X_i$ windows to form identity-aware representations $Z_i$. $Z_i$ contains the time-varying activity of each unit while also being informed of the unit's stable identity within one session. In matrix form:
\begin{equation}\label{eq: id}
    Z = X + E 
\end{equation}
where $Z_i$'s, $X_i$'s, $E_i$'s constitute rows of the $Z, X, E$ matrices.

We leverage the cross-attention mechanism to selectively aggregate information from population units and decode behavior outputs:
\begin{equation} \label{eq. cross-attn}
\begin{aligned}
    Y = \mathrm{CrossAttn}(Q, Z, Z) &= \mathrm{softmax} \left( \frac{Q K^\top}{\sqrt{d_k}} \right) V\\
\end{aligned}
\end{equation}
where $K = ZW_K, V = ZW_V \in \mathrm{R}^{N_s \times W}$ are projections of the identity-informed neural activity $Z$, and $Q \in \mathrm{R}^{B \times W}$ is a learnable matrix to query the behavior from Z. We use the standard cross-attention module with pre-normalization and feedforward layers.
\begin{prop}
    Cross-attention with identity-informed neural activity (Equation \ref{eq. cross-attn}) is invariant to the permutation of neural units, i.e.,
    \begin{equation}
        \mathrm{CrossAttn}(Q, Z, Z) = \mathrm{CrossAttn}(Q, P_RZ, P_RZ),
    \end{equation}
    where $P_R$ is the row permutation matrix. (See proof in Appendix).
\end{prop}
$E$ in Equations \ref{eq: id} and \ref{eq. cross-attn} can be understood as a special kind of positional embedding for attention mechanism, where $E$ is equivariant to the order of tokens (neural units), i.e., permuting the rows in $X$ also permutes the rows in $E$ accordingly. Hence, unlike traditional positional embeddings in the transformer literature where positional embeddings are fixed entities, our proposed $E$ is \emph{context dependent}. This context-dependent positional embedding enables cross-session generalization by design, as $E$ is stable for all samples within the same context (session), and can readily adapt in a gradient-free manner to new populations with arbitrary size and order.

After cross-attention, we project down $Y$ by a fully connected layer to a one-dimensional vector representing the predicted behavior covariates at the last timestep, based on which we compute the mean squared error (MSE) between the predicted and the ground truth behavior covariates. The IDEncoder and cross-attention module are trained in an end-to-end manner using this MSE objective.
\subsection{Encouraging model robustness to inconsistent population composition}
Neural population distributes its computation among many neural units, allowing us to effectively decode behavior even though we can only record neural activity with a limited number of electrodes. Leveraging this insight and in order to encourage model robustness to different compositions of neural membership across recording sessions, we employ \emph{dynamic channel dropout}, a novel technique to avoid overfitting to the population composition seen during training. Unlike classical population dropout methods \cite{keshtkaran2019enabling, keshtkaran2022large, ye2021representation} where only a fixed fraction of neurons/timesteps is zeroed-out during training, we randomly sample a dropout rate between $0$ and $1$ each training iteration and remove population units with the sampled dropout rate. With dynamic channel dropout, we not only encourage the model to be robust to the unit membership but also encourage it to be robust to the size of the population, leading to improved cross-session generalization (see Section \ref{section: ablation} and Figure \ref{fig 4: 2 ablation studies}).
\subsection{Gradient-free, few-shot cross-session adaptation in unseen neural populations with variable size and order}
The overall framework is depicted in Figure \ref{fig 2: algorithm}. We use labeled data from all training sessions to train all model parameters following the above pipeline. The model naturally digests populations of arbitrary size and order in all sessions without any need of session-specific alignment layer or fixed positional embeddings for neural units, hence having the potential to scale up to a large amount of data. When testing on a held-out session, we reuse the trained IDEncoder and only need a few \emph{unlabeled} calibration trials to infer identities of neural units in the test session, without the need of gradient descent updates to fine-tune session-specific alignment layers or unit/session embeddings. With these benefits, our proposed model removes the time and computation overhead usually required for re-calibrating neural decoders before each session, and facilitates its applicability in real-world iBCI settings where test-time labels are inherently unavailable. 
\section{Experiments}
\subsection{Datasets and evaluation metrics}
We evaluate our approach on three continuous motor decoding tasks from the Few-shot Algorithms for Consistent Neural Decoding (FALCON) Benchmark~\cite{karpowicz2024few}.
Specifically, we evaluate SPINT on the M1, M2, and H1 datasets. In M1, a monkey reached to, grasped, and manipulated an object in a variety of locations (4 possible objects, 8 locations), while neural activity was recorded from precentral gyrus and intramuscular electromyography (EMG) was recorded from 16 muscles \cite{rouse2015spatiotemporal, rouse2016spatiotemporal, rouse2016spatiotemporal2, rouse2018condition}. In M2, a monkey made finger movements to control a virtual hand and acquired cued target positions while neural activity from the precentral gyrus and 2-D actuator velocities were captured \cite{nason2021real}. In H1, a human subject attempted to reach and grasp with their right hand according to a cued motion for a 7-degree-of-freedom robotic arm control \cite{collinger2013high, wodlinger2014ten, flesher2021brain}.
Each dataset comprises multiple labeled held-in sessions used to train the decoder (spanning 4, 4, and 6 days for M1, M2, and H1, respectively), and multiple held-out sessions for model evaluation (spanning 3, 4, and 7 days for M1, M2, and H1, respectively). Each held-out session provides a few public calibration trials (with optional labels) used for decoder calibration, after which the decoder is evaluated on a private test split. Cross-session performance is quantified by the mean and standard deviation of $R^2$ between the predicted and ground truth behavior covariates across all held-out sessions. All evaluation results were obtained on the held-out private split by submitting models to the EvalAI platform \cite{evalai_falcon}.
\subsection{Baselines}
We compare SPINT with zero-shot (ZS) and few-shot unsupervised (FSU) baselines, since SPINT is the intersection of these two approaches. Similar to FSU approaches, SPINT makes use of a few unlabeled calibration samples in the held-out sessions; however, unlike conventional FSU approaches, SPINT does not require gradient updates for model parameters at test time, therefore bearing resemblance to ZS methods in terms of practical utility. We call this new class of model \emph{gradient-free few-shot unsupervised} (GF-FSU).

\textbf{ZS Wiener Filter and ZS RNN}: Wiener Filter is a linear model that predicts the current behavior as a weighted sum of previous timesteps \cite{wiener1964extrapolation}. In addition to the Wiener Filter, we also compare with a simple RNN baseline (implemented as an LSTM \cite{hochreiter1997long}).
The WF and RNN models were fitted using a single held-in session and evaluated zero-shot on the held-out sessions. 

\textbf{CycleGAN \cite{ma2023using}}: An FSU method where a Generative Adversarial Network (GAN) is trained using calibration data from a held-out session (day K) to transform day K's population activity to a form resembling activity from a held-in session (day 0), allowing decoders pretrained on day 0 to be reused on day K.

\textbf{NoMAD \cite{karpowicz2022stabilizing}}: Another FSU method where a dynamical model and a decoder are trained on day 0 to predict behavior from the inferred dynamics. Then on day K, an alignment network is trained to match the distribution of neural latent states to that of day 0, allowing the fixed model and decoder to transfer to day K. 

\textbf{Wiener Filter, RNN and Transformer Oracles (OR)}: We include the Wiener Filter, RNN, and NDT2 - a transformer for neural data \cite{ye2023neural}, trained on private held-out labeled data to serve as upper bounds for model performance.

\textbf{NDT2 Multi (FSS) \cite{ye2023neural}}: Similar to NDT2 Multi OR, but only trained on held-in and held-out few-shot calibration data with supervision. 
\subsection{SPINT outperforms zero-shot and few-shot unsupervised baselines on continuous motor decoding tasks}
\begin{table}[t]
    \centering
    \begin{tabular}{lcccc}
        & \textbf{Class} & \textbf{M1} & \textbf{M2} & \textbf{H1} \\
        \hline
        Wiener Filter (WF) & OR & $0.53 \pm 0.04$ & $0.26 \pm 0.03$ & $0.21 \pm 0.04$ \\
        RNN  & OR & $0.75 \pm 0.05$ & $0.56 \pm 0.04$ & $0.44 \pm 0.13$ \\
        NDT2 Multi \cite{ye2023neural}& OR & $0.78 \pm 0.04$ & $0.58 \pm 0.04$ & $0.63 \pm 0.08$ \\
        NDT2 Multi \cite{ye2023neural} & FSS & $0.59 \pm 0.07$ & $0.43 \pm 0.08$ & $0.52 \pm 0.04$ \\
        \hline
        WF & ZS & $0.34 \pm 0.06$ & $0.06 \pm 0.04$ & $0.16 \pm 0.03$ \\
        RNN & ZS & $-0.60 \pm 0.45$ & $-0.07 \pm 0.23$ & $0.09 \pm 0.18$ \\
        CycleGAN + WF \cite{ma2023using}  & FSU & $0.43 \pm 0.04$ & $0.22 \pm 0.06$ & $0.12 \pm 0.06$ \\
        NoMAD + WF \cite{karpowicz2022stabilizing} & FSU & $0.49 \pm 0.03$ & $0.20 \pm 0.10$ & $0.13 \pm 0.10$ \\
        \textbf{SPINT (Ours)} & GF-FSU & $\textbf{0.66} \pm 0.07$ & $\textbf{0.26} \pm 0.13$ & $\textbf{0.29} \pm 0.15 $\\
    \end{tabular}
    \vspace{0.5em}
    \caption{Performance comparison against oracles (OR), few-shot supervised (FSS), few-shot unsupervised (FSU), and zero-shot (ZS) methods. Our SPINT approach belongs to a special class which we termed gradient-free few-shot unsupervised (GF-FSU), where models perform adaptation based on few-shot unlabeled data but without any parameter updates at test time. Results are reported as mean $\pm$ standard deviation $R^2$ across held-out sessions.}
    \label{tab:methods_comparison}
\end{table}

We show in Table \ref{tab:methods_comparison} the performance of SPINT in comparison with ZS and FSU approaches. SPINT outperforms all ZS and FSU baselines across all three datasets, while requiring no retraining or fine-tuning of model parameters. Improvement is most prominent in M1, where the amount of training data is the largest ($\sim$$5\times$ more data than H1 and $\sim$$6\times$ more data than M2 in terms of recording time). Notably, SPINT surpasses Wiener Filter oracles in all datasets, which were trained with access to the private labeled data.
SPINT even outperforms the FSS method NDT2 Multi on M1 dataset while unlike NDT2, it does not require access to test-time labels or model parameter updates. As we focus on cross-session transferability, all our experimental results show the cross-session performance. We include comparison on within-session performance in the Appendix.

\begin{figure}
    \centering
    \includegraphics[width=1\linewidth]{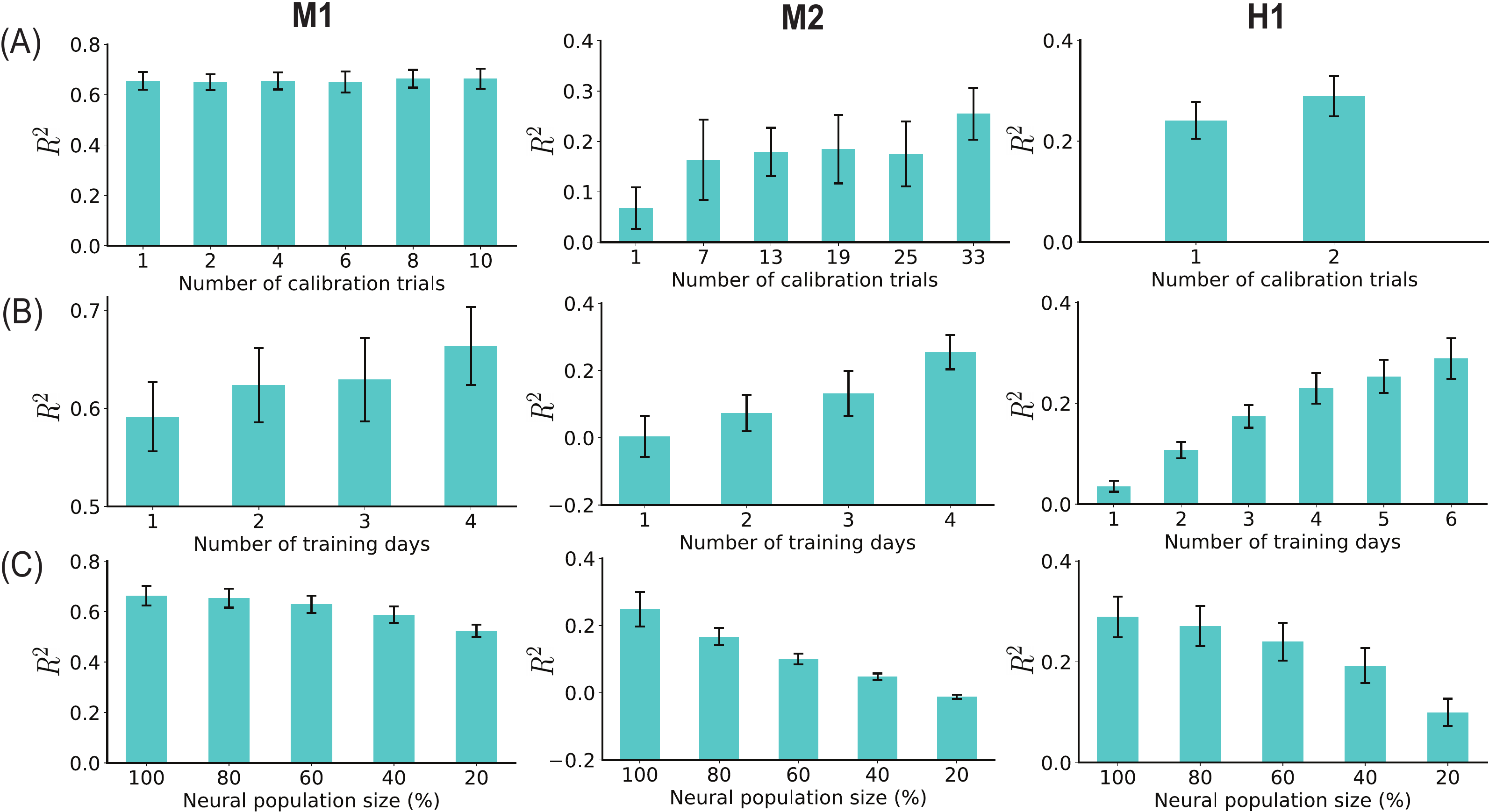}
    \caption{\textbf{Scaling analyses}. Cross-session performance of SPINT against number of calibration trials (A), training days (B), and population sizes (C) across M1, M2, and H1 datasets. 
    Bars represent mean $R^{2}$ across held-out sessions, whiskers represent standard error of the mean of $R^2$ across held-out sessions.}
    \label{figure 3: 3 main analysis}
\end{figure}

\subsection{SPINT requires only a minimal amount of unlabeled data for adaptation}
To gauge the data efficiency of our model at test time, we trained and tested the model with varying number of calibration trials used to infer the neural unit IDs. We show in Figure \ref{figure 3: 3 main analysis}(A) that SPINT could achieve reasonable cross-session generalization with a small number of few-shot trials. In M1 dataset, the model could even achieve similar performance as the best model (which uses all available calibration trials) with only one single trial. This study demonstrates the practical utility of SPINT in online iBCI, relieving the burden of data collection and label collection on users at test time. 

\subsection{SPINT performance scales well with the amount of training data}
Thanks to the flexible permutation-invariant transformer network and the context-dependent positional embeddings, SPINT can ingest populations with arbitrary sizes and orders. These design choices give SPINT the ability to scale naturally with large amounts of training data. We demonstrate this scaling ability in Figure \ref{figure 3: 3 main analysis}(B), where we observe a clear trend in cross-session performance when we use data from more held-in days to train the model, with the best performance achieved when using all available training data on each dataset. 
This ability suggests potentials of SPINT as a large-scale pretrained model for iBCI when trained on larger datasets beyond the FALCON benchmark.

\subsection{SPINT is robust to variable population composition}
Our proposed dynamic channel dropout encourages robustness of SPINT to variable input population size and membership. We test this robustness by training SPINT on the full held-in populations with dynamic channel dropout and evaluating on variable-sized held-out populations (Figure \ref{figure 3: 3 main analysis}C). At each evaluation batch, we randomly sample a subset of the original population and measure the $R^2$ obtained when the model makes predictions based on this limited subset. We observe robust performance in M1 with reasonable performance drop when the population gets increasingly smaller, with the model still achieving a mean held-out $R^2$ of $0.52$ when only $20\%$ of the original population remains, outperforming other ZS and FSU baselines with the full population.

\subsection{SPINT maintains low-latency inference for iBCI systems}
\begin{table}[t]
    \centering
    \begin{tabular}{lcccc}
        & \textbf{Class} & \textbf{M1} & \textbf{M2} & \textbf{H1} \\
        \hline
        Wiener Filter (WF) & OR & $0.06$ & $0.08$ & $0.14$ \\
        RNN  & OR & $0.04$ & $0.04$ & $0.08$ \\
        NDT2 Multi & OR & $0.15$ & $0.10$ & $2.29$ \\
        NDT2 Multi & FSS & $0.13$ & $0.10$ & $0.30$ \\
        \hline
        WF & ZS & $0.06$ & $0.08$ & $0.15$ \\
        RNN & ZS & $0.03$ & $0.01$ & $0.02$ \\
        CycleGAN + WF & FSU & $0.07$ & $0.09$ & $0.16$ \\
        NoMAD + WF & FSU & $0.99$ & $0.91$ & $1.03$ \\
        \textbf{SPINT (Ours)} & GF-FSU & $0.13$ & $0.13$ & $0.14$\\
    \end{tabular}
    \vspace{0.5em}
    \caption{Inference latency of SPINT against oracles (OR), few-shot supervised (FSS), few-shot unsupervised (FSU) and zero-shot (ZS) methods on held-out sessions (lower is better).}
    \label{tab:latency}
\end{table}

A critical consideration in iBCI system deployment is the ability of the system to perform behavior decoding in real time. We designed SPINT with this consideration in mind, using only one layer of cross-attention and two three-layer fully connected networks for IDEncoder. In Table \ref{tab:latency}, we report the latency achieved by SPINT as compared to other methods. Latency is defined as the amount of time a method requires to process the evaluation data divided by the duration of the evaluation data \cite{evalai_falcon}. The ratio less than $1$ signifies the approximation to real-time iBCI inference. SPINT achieves $0.13$ latency on M1 and M2, and $0.14$ latency on H1, matching or outperforming transformer baselines, while being significantly below $1$. In practice, SPINT could be potentially faster in terms of deployment time, as it eliminates the need for an explicit alignment step required by conventional iBCI systems.

\subsection{Ablation Study}\label{section: ablation}
\begin{figure}
    \centering
    \includegraphics[width=1.0\linewidth]{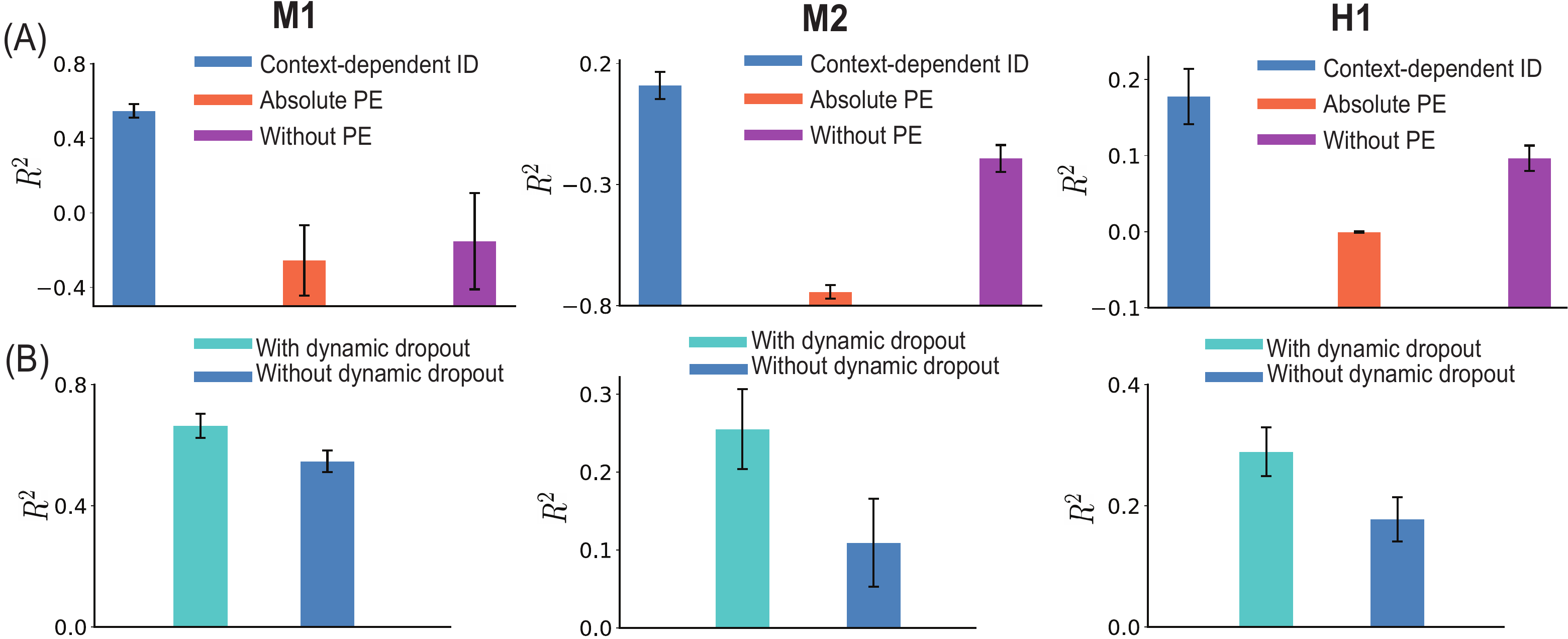}
    \caption{\textbf{Ablation Study}. Analyses showing the critical roles of our proposed context-dependent ID against fixed positional embeddings (PE) and no positional embeddings (A), our dynamic channel dropout against no dynamic channel dropout (B). Results are shown across M1, M2, and H1 datasets.
    Bars represent mean $R^{2}$ across held-out sessions, whiskers represent standard error of the mean of $R^2$ across held-out sessions.}
    \label{fig 4: 2 ablation studies}
\end{figure}

We perform ablation studies to demonstrate the benefits of our context-dependent positional embeddings and dynamic channel dropout techniques. In Figure \ref{fig 4: 2 ablation studies}A, we compare our context-dependent positional embeddings with fixed (absolute) positional embeddings used in the vanilla transformer \cite{vaswani2017attention}, and with no positional embeddings. The conventional fixed positional embeddings break the permutation-invariance property of the cross-attention mechanism, thus are not able to generalize to populations with different compositions in held-out sessions. With no positional embeddings, the model is permutation-invariant by design; however, the loss of information about neural unit functional identities hinders the model's ability to decode the behavior these units encode. We achieve the best of both worlds by our proposed context-dependent positional embeddings, being both permutation-invariant while retaining neural identities for behavior decoding.

To demonstrate the effectiveness of our proposed dynamic channel dropout technique, we compare the cross-session performance of SPINT with dynamic channel dropout and without dynamic channel dropout. We show in Figure \ref{fig 4: 2 ablation studies}B that dynamic channel dropout serves as an effective regularization technique by preventing the model from overfitting to the population composition in training sessions.

\section{Discussion}
In this work, we introduce SPINT, a permutation-invariant transformer designed for cross-session intracortical motor decoding. SPINT features a context-dependent neural ID embeddings that dynamically infers unit identities, allowing the model to handle unordered, variable-sized populations across sessions. To enhance robustness to population compositions, we proposed dynamic channel dropout, which simulates unit instability during training. These components enable SPINT to adapt to new sessions using only a small amount of unlabeled calibration data, with no need for test-time labels or gradient updates.
We demonstrate cross-session generalization ability of SPINT on three continuous motor decoding tasks from the FALCON Benchmark, where it consistently outperforms existing zero-shot and few-shot unsupervised baselines, even surpassing few-shot supervised and oracle models in some instances. SPINT's light-weight design also enables low inference latency, making it well-suited for real-world iBCI applications. 

Our study represents an initial attempt at a flexible, gradient-free framework for consistent behavioral decoding and opens up several promising avenues for future research. Within the framework of the FALCON Benchmark, we demonstrated that our approach scales effectively with increasing amount of training data, albeit from a single subject and behavior task. Exploring the applicability of our approach under more diverse settings (cross-task, cross-region, cross-subject) is an important direction for future research \cite{ye2021representation, azabou2023unified, zhang2025neural}. On the technical side, our end-to-end training for the unit identifier together with the rest of the network ties the identification of the neural identities to task-specific behavioral decoding and thus requires access to labels to train the unit identifier, which might be limited in real-world iBCI settings. Disentangling the training of the unit identifier with behavioral decoding by means of self-supervised or contrastive learning approaches can potentially help alleviate the reliance on behavior labels for this stage. 

Lastly, although our study shows robust and low-latency behavioral decoding \emph{in silico}, its performance remains to be evaluated \emph{in vivo}, where sensory feedback from closed-loop control may induce modulation in the neural activity inputs and hardware-specific constraints could affect the online inference latency \cite{orsborn2014closed}. To this end, our preliminary assessment represents an initial step toward a light-weight, plug-and-play decoding framework for long-term iBCI systems.

\bibliographystyle{unsrt}
\bibliography{reference}
\beginappendix
\section{Appendix}
\subsection{Within-session performance comparison}
We include the within-session performance comparison between SPINT and baselines in Table \ref{tab:methods_comparison_heldin}. This table is similar to Table 1 in the main paper, but with metrics obtained on EvalAI's private splits within the held-in sessions. As observed from the table, SPINT also consistently outperforms ZS and FSU baselines on the held-in splits.

\begin{table}[h]
    \centering
    \begin{tabular}{lcccc}
        & \textbf{Class} & \textbf{M1} & \textbf{M2} & \textbf{H1} \\
        \hline
        Wiener Filter (WF) & OR & $0.54 \pm 0.01$ & $0.27 \pm 0.02$ & $0.24 \pm 0.02$ \\
        RNN  & OR & $0.75 \pm 0.03$ & $0.59 \pm 0.07$ & $0.51 \pm 0.09$ \\
        NDT2 Multi \cite{ye2023neural}& OR & $0.77 \pm 0.03$ & $0.62 \pm 0.03$ & $0.68 \pm 0.05$ \\
        NDT2 Multi \cite{ye2023neural} & FSS & $0.77 \pm 0.03$ & $0.63 \pm 0.03$ & $0.62 \pm 0.04$ \\
        \hline
        WF & ZS & $0.46 \pm 0.06$ & $0.15 \pm 0.07$ & $0.20 \pm 0.04$ \\
        RNN & ZS & $0.52 \pm 0.15$ & $0.20 \pm 0.29$ & $0.31 \pm 0.13$ \\
        CycleGAN + WF \cite{ma2023using}  & FSU & $0.61 \pm 0.02$ & $0.32 \pm 0.03$ & $0.15 \pm 0.04$ \\
        NoMAD + WF \cite{karpowicz2022stabilizing} & FSU & $0.64 \pm 0.01$ & $0.35 \pm 0.05$ & $0.21 \pm 0.06$ \\
        \textbf{SPINT (Ours)} & GF-FSU & $\textbf{0.77} \pm 0.02$ & $\textbf{0.59} \pm 0.01$ & $\textbf{0.47} \pm 0.06 $\\
    \end{tabular}
    \vspace{0.5em}
    \caption{Within-session performance comparison against oracles (OR), few-shot supervised (FSS), few-shot unsupervised (FSU), and zero-shot (ZS) methods. Our SPINT approach belongs to a special class which we termed Gradient-Free Few-Shot Unsupervised (GF-FSU), where models perform adaptation based on few-shot unlabeled data but \emph{without} any parameter updates at test time. Results are reported as mean $\pm$ standard deviation $R^2$ across held-in sessions, achieved on EvalAI private held-in splits.}
    \label{tab:methods_comparison_heldin}
\end{table}

\subsection{Proof of SPINT's permutation-invariance}
Let $P_R$, $P_C$ be the row and column permutation matrices of the same permutation $\pi$ ($P_C = P_R^\top = P_R^{-1}$ and $P_C P_R = I$). Also let $X' = P_R X$ and $(X^C)' = P_R X^C$ be the row-permuted neural windows and row-permuted calibration trials.

Since the ID embedding of each neural unit $i$ is computed individually from the set of calibration trials for that unit:
\begin{equation}
    E_i = \mathrm{IDEncoder}(X_{i}^{C}) = \psi(pool(\phi(X_{i}^{C}))),
\end{equation}
permuting the neural units in the original population (neural windows $X$ or calibration trials $X^C$) will permute the embedding matrix $E$ in the exact same order, i.e., $E' = P_R E$.

It follows that:
\begin{equation}
    Z' = X' + E' = P_R X + P_R E = P_R (X + E) = P_R Z
\end{equation}
In other words, $Z$ is equivariant to the permutation of neural units. 

Cross-attention performed on $Z'$ then becomes:
\begin{equation}\label{eq: cross-attn}
\begin{aligned}
    \mathrm{CrossAttn}(Q, Z', Z') &= \mathrm{CrossAttn}(Q, P_R Z, P_R Z) \\
    &= \mathrm{softmax} \left( \frac{Q W_K^\top Z^\top P_R^\top}{\sqrt{d_k}} \right) P_R Z W_V\\
    &= \mathrm{softmax} \left( \frac{Q W_K^\top Z^\top P_C}{\sqrt{d_k}} \right) P_R Z W_V\\
    &= \mathrm{softmax} \left( \frac{Q W_K^\top Z^\top }{\sqrt{d_k}} \right) P_C P_R Z W_V\\
    &= \mathrm{softmax} \left( \frac{Q W_K^\top Z^\top }{\sqrt{d_k}} \right) Z W_V\\ 
    &= \mathrm{CrossAttn}(Q, Z, Z)
\end{aligned}
\end{equation}
where $\mathrm{softmax} \left( \frac{Q W_K^\top Z^\top P_C}{\sqrt{d_k}} \right) = \mathrm{softmax} \left( \frac{Q W_K^\top Z^\top}{\sqrt{d_k}} \right) P_C$ because an element is always normalized with the same group of elements in the same row regardless of whether column permutation is performed before or after $\mathrm{softmax}$. 

Equation \ref{eq: cross-attn} concludes Proposition 1 in the main paper.

We note that multi-layer perceptron (MLP), layer normalization, and residual connection are applied row-wise and hence do not affect the overall permutation-invariance property of our SPINT framework. 

\subsection{Correlation of attention scores and firing statistics}
We ask whether the attention scores SPINT assigns for each neural unit are correlated with its firing statistics. To answer this question, in each held-out calibration window, we measure the average attention scores over $B$ behavior covariates, and its firing statistics (mean/standard deviation) over the held-out calibration trials, then calculate the Pearson's correlation between these two quantities using all held-out calibration windows. We show the results in Table \ref{tab:attn_scores_fr_correlation}.

We observe that the attention scores correlate moderately with the mean and the standard deviation of the neural unit's firing rates, with higher correlation for the standard deviation than the mean, suggesting that SPINT might be extracting neural units that are active (having high mean firing rates) and behaviorally relevant (having high variance throughout the calibration periods where behavior is varied) to pay attention to in behavioral decoding.
\begin{table}[h]
    \centering
    \begin{tabular}{lccc}
        & \textbf{M1} & \textbf{M2} & \textbf{H1} \\
        \hline
        $\rho(\text{attention scores, mean firing rates})$ & $0.33 \pm 0.16$ & $0.76 \pm 0.03$ & $0.51 \pm 0.04$ \\
        $\rho(\text{attention scores, standard deviation of firing rates})$ & $0.45 \pm 0.16$ & $0.87 \pm 0.02$ & $0.57 \pm 0.03$ \\
    \end{tabular}
    \vspace{0.5em}
    \caption{Pearson's correlation between attention scores for each neural unit and that unit mean/standard deviation of firing rates during the held-out calibration periods. Results are reported as the mean correlation $\pm$ standard deviation across held-out sessions. All $p$-values are less than $0.05$.}
    \label{tab:attn_scores_fr_correlation}
\end{table}

\subsection{Implementation details}
\subsubsection{Data preprocessing}
For neural activity, we use the binned spike count obtained by unit threshold crossing with the standard bin size of 20ms as set forth by the FALCON Benchmark. We follow FALCON's continuous decoding setup for all three M1, M2, and H1 datasets, where rather than decoding trialized behavior from the trialized neural activity (often performed in a non-causal manner), we decode behavior at the last step of a neural activity window, mimicking the online, causal iBCI decoding. To construct the length-$W$ neural window at the beginning of each session, we pre-pad the session neural time series with $(W-1)$ zeros. We discard the windows whose last time step belongs to a non-evaluated period as defined by FALCON, e.g., inter-trial periods where there is no registered kinematics.

Our IDEncoder infers neural unit identity from trialized calibration trials. As calibration trials vary in length, we interpolate all calibration trials to the same length $T$, where $T=100$ for M2 and $T=1024$ for M1 and H1. We use the Python library \texttt{scipy.interpolate.interp1d} with a cubic spline for interpolation. Note that we only perform interpolation for neural calibration trials to synchronize their trial lengths. We still use the raw spike counts for the neural windows, conforming with the continuous decoding setup.

\subsubsection{Behavior output scaling}
For M2 and H1, since values of behavior covariates are relatively small, during training we scale the network behavior predictions by a factor of $0.2$ and $0.05$ for M2 and H1, respectively, effectively asking the model to predict $5\times$ and $20\times$ the original behavior values. The MSE loss and $R^2$ metrics are computed between the scaled predicted outputs and the original ground truth values.

\subsubsection{Inferring neural unit identity}
We follow the permutation-invariant framework in \cite{zaheer2017deep} for inferring identity $E_i$ of neural unit $i$:
\begin{equation}
    E_i = \mathrm{IDEncoder}(X_{i}^{C}) 
    = \mathrm{MLP_2}(\frac{1}{M}\sum_{j=1}^{M}(\mathrm{MLP_1}(X_{i}^{C_j})))
\end{equation}
where $M$ is the number of calibration trials, $X_i^{C_j}$ is the neural activity of the $j^{th}$ calibration trial of neural unit $i$, $\mathrm{MLP_1}$ and $\mathrm{MLP_2}$ are two 3-layer fully connected networks. $\mathrm{MLP_1}$ projects the length-$T$ trials to a hidden dimension $H$, and $\mathrm{MLP_2}$ projects the length-$H$ hidden features to length-$W$ neural unit identity output. 

\subsubsection{Behavioral decoding by cross-attention}
After neural identity for all units $E$ is inferred, we add it to the neural window input $X$ to form the identity-aware neural activity $Z$, i.e., $Z=X+E$. We then use the cross-attention mechanism in the latent space to decode last step behavior covariates. Specifically:
\begin{equation}
    Z_{in} = \mathrm{MLP_{in}}(Z)
\end{equation}
\begin{equation}
    \tilde{Z} = Z_{in} + \mathrm{CrossAttn}(Q, \mathrm{LayerNorm}(Z_{in}), \mathrm{LayerNorm}(Z_{in}))
\end{equation}
\begin{equation}
    Z_{out} = \tilde{Z} + \mathrm{MLP_{attn}}(\mathrm{LayerNorm}(\tilde{Z}))
\end{equation}
\begin{equation}
    Y = \mathrm{MLP_{out}}(Z_{out})
\end{equation}

\subsubsection{Hyperparameters}
We include the notable hyperparameters used to optimize SPINT in Table \ref{tab:hyperparams}. We train and evaluate models for each M1, M2, and H1 dataset separately. We train the models using all available held-in sessions and evaluate on all available held-out sessions. We use Adam optimizer \cite{kingma2014adam} for all training. 

\begin{table}[h]
    \centering
    \begin{tabular}{lccc}
        & \textbf{M1} & \textbf{M2} & \textbf{H1} \\
        \hline
        Batch size & $32$ & $32$ & $32$ \\
        Window size & $100$ & $50$ & $700$ \\
        Max trial length & $1024$ & $100$ & $1024$ \\
        Number of IDEncoder layers & $3,3$ & $3,3$ & $3,3$ \\
        Number of cross attention layers & $1$ & $1$ & $1$ \\
        Hidden dimension & $1024$ & $512$ & $1024$ \\
        Behavior scaling factor & $1$ & $0.2$ & $0.05$ \\
        Learning rate & $1\mathrm{e}{-5}$ & $5\mathrm{e}{-5}$ & $1\mathrm{e}{-5}$ \\
    \end{tabular}
    \vspace{0.5em}
    \caption{Hyperparameters used to train SPINT on the M1, M2, and H1 datasets.}
    \label{tab:hyperparams}
\end{table}

\subsubsection{Computational resources}
SPINT was trained using a single A40 GPU, consuming less than 2GB of GPU memory with batch size of 32 and taking around 12 hours, 5 hours, and 8 hours to finish 50 training epochs for M1, M2, and H1, respectively. We select checkpoints for evaluation at epoch 50 in all M1, M2, and H1 datasets.
\end{document}